\documentclass[cits]{PoS}

\title{Theoretical progress on the $\mathbf{V_{us}}$ determination from $\mathbf{\tau}$ decays}
\ShortTitle{Theoretical progress on the $V_{us}$ determination from $\mathbf{\tau}$ decays}

\author{Elvira~G\'amiz,$^a$ Matthias~Jamin,$^{bc}$
 \speaker{Antonio Pich},$^{d}$ Joaquim~Prades$^{ef}$ and Felix~Schwab$^{c}$\\
\llap{$^a$}Department of Physics, University of Illinois,
Urbana IL 61801, USA\\
\llap{$^b$}ICREA\\
\llap{$^c$}IFAE, Departament de F\'{\i}sica Te\`orica,
Universitat Auton\`oma de Barcelona,\\ E-08193 Bellaterra, Barcelona, Spain\\
\llap{$^d$}Departament de F\'{\i}sica Te\`orica, IFIC,
Universitat de Val\`encia--CSIC, \\  Apt. de Correus 22085, E-46071
Val\`encia, Spain\\
\llap{$^e$}Theory Unit, Physics Department, CERN, CH-1211 Gen\`eve 23,
Switzerland\\
\llap{$^f$}CAFPE and Departamento de
F\'isica Te\'orica y del Cosmos, Universidad de Granada,\\
Campus de Fuente Nueva, E-18002 Granada, Spain\\
E-mail: \email{megamiz@uiuc.edu}, \email{jamin@ifae.es},
\email{antonio.pich@ific.uv.es}, \email{prades@ugr.es}, \email{schwab@ifae.es}}

\abstract{A very precise determination of $V_{us}$ can be obtained from
the semi-inclusive hadronic decay width of the $\tau$ lepton into final
states with strangeness. The ratio of the Cabibbo-suppressed and Cabibbo-allowed
$\tau$ decay widths directly measures $(V_{us}/V_{ud})^2$, up to very small
SU(3)-breaking corrections which can be theoretically estimated with the
needed accuracy.
Together with previous LEP and CLEO data,
the recent measurements by Babar and Belle of some Cabibbo-suppressed
$\tau$ decays imply
$V_{us}= 0.2165\pm 0.0026_{\mathrm{exp}}\pm 0.0005_{\mathrm{th}}$, which
is already competitive with the standard extraction from $K_{l3}$ decays.
The uncertainty is largely dominated by
experimental errors and should be easily reduced with the high statistics
of the B factories, providing the most accurate determination
of this parameter. A 1\% experimental precision on the Cabibbo-suppressed
$\tau$ decay width would translate into a 0.6\% uncertainty on $V_{us}$.}

\FullConference{Kaon International Conference\\
May 21-25, 2007\\
Laboratori Nazionali di Frascati dell'INFN}

\begin{document}

\section{The hadronic $\mathbf{\tau}$ decay width}

The hadronic decays of the $\tau$ lepton provide a very clean laboratory
to perform precise tests of the Standard Model \cite{tau06}.
The inclusive character of the total $\tau$ hadronic width renders
possible an accurate calculation of the ratio
\cite{BR:88,NP:88,BNP:92,LDP:92a,QCD:94}
%
\begin{equation}\label{eq:r_tau_def}
 R_\tau \,\equiv\, { \Gamma [\tau^- \to \nu_\tau
 \,\mathrm{hadrons}\, (\gamma)] \over \Gamma [\tau^- \to \nu_\tau e^-
 {\bar \nu}_e (\gamma)] }\, = \, R_{\tau,V} + R_{\tau,A} + R_{\tau,S}\, ,
\end{equation}
using analyticity constraints and the Operator Product Expansion.
One can separately compute the contributions associated with
specific quark currents:
$R_{\tau,V}$ and $R_{\tau,A}$ correspond to the Cabibbo-allowed
decays through the vector and axial-vector currents, while
$R_{\tau,S}$ contains the remaining Cabibbo-suppressed
contributions.

The theoretical prediction for $R_{\tau,V+A}$ can be expressed as
\cite{BNP:92}
\begin{equation}\label{eq:Rv+a}
 R_{\tau,V+A} \, =\, N_C\, |V_{ud}|^2\, S_{\mathrm{EW}} \left\{ 1 +
 \delta_{\mathrm{P}} + \delta_{\mathrm{NP}} \right\} ,
\end{equation}
where $N_C=3$ is the number of quark colours
and $S_{\mathrm{EW}}=1.0201\pm 0.0003$ contains the
electroweak radiative corrections \cite{MS:88,BL:90,ER:02}.
The dominant correction ($\sim 20\%$) is the perturbative QCD
contribution $\delta_{\mathrm{P}}$, which is fully known to
$O(\alpha_s^3)$ \cite{BNP:92} and includes a resummation of the most
important higher-order effects \cite{LDP:92a,PI:92}.

Non-perturbative contributions are suppressed by six powers of the
$\tau$ mass \cite{BNP:92} and, therefore, are very small. Their
numerical size has been determined from the invariant-mass
distribution of the final hadrons in $\tau$ decay, through the study
of weighted integrals \cite{LDP:92b},
\begin{equation}
 R_{\tau}^{kl} \,\equiv\, \int_0^{m_\tau^2} ds\, \left(1 - {s\over
 m_\tau^2}\right)^k\, \left({s\over m_\tau^2}\right)^l\, {d
 R_{\tau}\over ds} \, ,
\end{equation}
which can be calculated theoretically in the same way as $R_{\tau}$.
The predicted suppression \cite{BNP:92} of the non-perturbative
corrections has been confirmed by ALEPH \cite{ALEPH:05}, CLEO
\cite{CLEO:95} and OPAL \cite{OPAL:98}. The most recent analysis
\cite{ALEPH:05} gives
\begin{equation}\label{eq:del_np}
 \delta_{\mathrm{NP}} \, =\, -0.0043\pm 0.0019 \, .
\end{equation}

The QCD prediction for $R_{\tau,V+A}$ is then completely dominated
by the perturbative contribution; non-perturbative effects being
smaller than the perturbative uncertainties from uncalculated
higher-order corrections. The result turns out to be very sensitive
to the value of $\alpha_s(m_\tau)$, allowing for an accurate
determination of the fundamental QCD coupling \cite{NP:88,BNP:92}.
The experimental measurement $R_{\tau,V+A}= 3.471\pm0.011$ implies
\cite{DHZ:05}
\begin{equation}\label{eq:alpha}
 \alpha_s(m_\tau)  \,=\,  0.345\pm
 0.004_{\mathrm{exp}}\pm 0.009_{\mathrm{th}} \, .
\end{equation}

The strong coupling measured at the $\tau$ mass scale is
significantly larger than the values obtained at higher energies.
From the hadronic decays of the $Z$, one gets $\alpha_s(M_Z) =
0.1186\pm 0.0027$ \cite{LEPEWWG}, which differs from the $\tau$
decay measurement by more than twenty standard deviations. After
evolution up to the scale $M_Z$ \cite{Rodrigo:1998zd}, the strong
coupling constant in (\ref{eq:alpha}) decreases to \cite{DHZ:05}
\begin{equation}\label{eq:alpha_z}
 \alpha_s(M_Z)  \, =\,  0.1215\pm 0.0012 \, ,
\end{equation}
in excellent agreement with the direct measurements at the $Z$ peak
and with a similar accuracy. The comparison of these two
determinations of $\alpha_s$ in two extreme energy regimes, $m_\tau$
and $M_Z$, provides a beautiful test of the predicted running of the
QCD coupling; i.e., a very significant experimental verification of
{\it asymptotic freedom}.

\section{Cabibbo-suppressed $\mathbf{\tau}$ decay width}

The separate measurement of the $|\Delta S|=0$ and $|\Delta S|=1$ \
$\tau$ decay widths allows us to pin down the SU(3)-breaking effect
induced by the strange quark mass
\cite{Davier,PP:99,ChDGHPP:01,ChKP:98,KKP:01,MW:06,KM:00,MA:98,GJPPS:05,BChK:05},
through the differences
\begin{equation}
 \delta R_\tau^{kl}  \,\equiv\,
 {R_{\tau,V+A}^{kl}\over |V_{ud}|^2} - {R_{\tau,S}^{kl}\over |V_{us}|^2}\, .
\end{equation}
Since QCD is flavour blind, these quantities vanish in the SU(3) limit.
The only non-zero contributions are proportional to powers of
the quark mass-squared difference $m_s^2(m_\tau)-m_d^2(m_\tau)$ or to vacuum expectation
values of SU(3)-breaking operators such as $\langle\delta O_4\rangle
\equiv \langle 0|m_s\bar s s - m_d\bar d d|0\rangle \approx (-1.4\pm 0.4)
\cdot 10^{-3}\; \mathrm{GeV}^4$ \cite{PP:99,GJPPS:05}. The dimensions of these operators
are compensated by corresponding powers of $m_\tau^2$, which implies a strong
suppression of $\delta R_\tau^{kl}$ \cite{PP:99}:
\begin{equation}\label{eq:dRtau}
 \delta R_\tau^{kl}\,\approx\,  24\, S_{\mathrm{EW}}\,\left\{ {m_s^2(m_\tau)\over m_\tau^2} \,
 \left( 1-\epsilon_d^2\right)\,\Delta_{kl}(\alpha_s)
 - 2\pi^2\, {\delta O_4\over m_\tau^4} \, Q_{kl}(\alpha_s)\right\}\, ,
\end{equation}
where $\epsilon_d\equiv m_d/m_s = 0.053\pm 0.002$ \cite{LE:96}.
The perturbative QCD corrections $\Delta_{kl}(\alpha_s)$ and
$Q_{kl}(\alpha_s)$ are known to $O(\alpha_s^3)$ and $O(\alpha_s^2)$,
respectively \cite{PP:99,BChK:05}.

The moments $\delta R_\tau^{k0}$ ($k=0,1,2,3,4$) have been measured by
ALEPH \cite{ALEPHms} and OPAL \cite{OPALms}. In spite of the large
experimental uncertainties, the corresponding QCD
analysis has allowed to perform a rather competitive determination
of the strange quark mass \cite{ChDGHPP:01,GJPPS:05}.
However, the extracted value depends sensitively on the
modulus of the Cabibbo--Kobayashi--Maskawa matrix element
$|V_{us}|$, because the small differences $\delta R_\tau^{kl}$ result from
a strong cancelation between two nearly-equal quantities.
It appears then natural to turn things around and, with
an input for $m_s$ obtained from other sources, to actually
determine $|V_{us}|$ \cite{GJPPS:05}. The most sensitive moment is
the unweighted difference of decay widths $\delta R_\tau\equiv \delta R_\tau^{00}$:
\begin{equation}\label{eq:Vus_formula}
 |V_{us}|^2 \,=\, \frac{R_{\tau,S}}{\frac{R_{\tau,V+A}}{|V_{ud}|^2}-\delta
 R_{\tau,\mathrm{th}}} \, .
\end{equation}

The SU(3)-breaking quantity $\delta R_\tau\sim 0.25$
is one order of magnitude smaller than the ratio
$R_{\tau,V+A}/|V_{ud}|^2 = 3.661\pm 0.012$, where we have taken for $V_{ud}$
the PDG advocated value $|V_{ud}| = 0.97377\pm 0.00027$ \cite{PDG}.
Therefore, to a first approximation $V_{us}$ can be directly obtained
from experimental measurements, without any theoretical input.
With $R_{\tau,S}=0.1686\pm 0.0047$ \cite{DHZ:05}, one gets in the
SU(3) limit:
\begin{equation}
 \delta R_\tau =0\qquad\longrightarrow\qquad
 |V_{us}|^{\mathrm{SU(3)}} = 0.215\pm 0.003\, .
\end{equation}
This rather remarkable determination is only slightly shifted
by the small SU(3)-breaking corrections. For instance, taking
$\delta R_\tau\approx 0.25$ increases the result to
$|V_{us}|\approx 0.222$. Thus, an estimate of
$\delta R_{\tau,\mathrm{th}}$ with an accuracy of around
10\% translates into a final theoretical uncertainty for $|V_{us}|$
of only 0.4\% ($\pm 0.0008$). The final precision on the $\tau$
determination of $|V_{us}|$ is then a purely experimental issue,
in contrast to the standard extraction from $K_{l3}$ which is
already limited by theoretical errors.

\section{Theoretical evaluation of $\mathbf{\delta R_{\tau}}$}

The theoretical analysis of $R_\tau$  \cite{BR:88,NP:88,BNP:92}
involves the two-point correlation functions
\begin{equation}
 \Pi^{\mu\nu}_{ij,\mathcal{J}}(q) \,\equiv\, i \int {\rm d}^4 x \;
 e^{i q \cdot x} \,\langle 0 | T [ \mathcal{J}_{ij}^\mu(x)
 \mathcal{J}^{\nu\dagger}_{ij}(0) ]| 0 \rangle\, =\,
 \left[q^\mu q^\nu - q^2 g^{\mu\nu}\right]\Pi^T_{ij,\mathcal{J}}(q^2)
 + q^\mu q^\nu\, \Pi^L_{ij,\mathcal{J}}(q^2)
\end{equation}
of vector,
$\mathcal{J}^\mu_{ij}=V^\mu_{ij}\equiv\bar q_j\gamma^\mu q_i$,
and axial-vector,
$\mathcal{J}^\mu_{ij}=A^\mu_{ij}\equiv\bar q_j\gamma^\mu\gamma_5 q_i$,
quark currents ($i,j=u,d,s$).
The invariant-mass distribution of the final
hadrons in the $\tau$ decay is proportional to the
imaginary parts of the correlators
$\Pi^{T/L}_{ij,\mathcal{J}}(q^2)$, where the superscript in the
transverse and longitudinal components denotes the corresponding
angular momentum $J=1$ ($T$) and $J=0$ ($L$).
Employing the analytic properties of these correlators
one can express $R_\tau$  as a contour integral running counter-clockwise
around the circle $|s|=m_\tau^2$ in the complex s-plane:
\begin{eqnarray}\label{eq:contour}
R_\tau & = &
12\pi\int\limits_0^{m_\tau^2} \frac{ds}{m_\tau^2}\;\biggl(
1-\frac{s}{m_\tau^2}\biggr)^2\, \biggl[\,\biggl(1+2\frac{s}{m_\tau^2}\biggr)\,
\mathrm{Im}\,\Pi^T(s)\, +\,\mathrm{Im}\,\Pi^L(s)\,\biggr]
\nonumber\\
&=& -\,i \pi \oint_{|s|=m_\tau^2}
\frac{{\rm d} s}{s} \; \left[1-\frac{s}{m_\tau^2}\right]^3 \,
 \left\{ 3 \left[1+\frac{s}{m_\tau^2}
\right] D^{L+T}(s)\, +\, 4 \, D^L(s) \right\} \, .
\end{eqnarray}
We have used integration by parts to rewrite $R_\tau$ in terms of
the logarithmic derivatives
\begin{equation}
D^{L+T}(s) \,\equiv\, -\,s \, \frac{{\rm d}}{{\rm d } s }
\, \Pi^{L+T}(s) \, , \qquad\qquad
D^{L}(s) \,\equiv\, \frac{s}{m_\tau^2} \, \frac{{\rm d} }{{\rm d } s}
\, \left[ s\, \Pi^{L}(s) \right] \, \, ,
\end{equation}
where the relevant combination of two-point correlation functions is given by
\begin{equation}
\Pi^J(s) \,\equiv\, |V_{ud}|^2 \left\{ \Pi^J_{ud,V}(s) +
\Pi^J_{ud,A}(s) \right\} + |V_{us}|^2 \left\{ \Pi^J_{us,V}(s) + \Pi^J_{us,A}(s) \right\}
\, .
\end{equation}
The two terms proportional to $|V_{ud}|^2$ contribute to $R_{\tau, V}$ and $R_{\tau, A}$,
respectively, while $R_{\tau, S}$  contains the remaining contributions
proportional to $|V_{us}|^2$.

At large enough Euclidean $Q^2\equiv -s$, both $\Pi^{L+T}(Q^2)$ and
$\Pi^{L}(Q^2)$ can be computed within QCD, using well-established
operator product expansion techniques. The result is organised in a
series of local gauge-invariant operators of increasing dimension, times
the appropriate inverse powers of $Q^2$. Performing the complex integration
(\ref{eq:contour}), one can then express $R_\tau$ as an expansion in inverse
powers of $m_\tau^2$ \cite{BNP:92}.
The perturbative correction $\delta_P$ in eq.~(\ref{eq:Rv+a}) corresponds to
the dimension-zero contributions.  The dominant SU(3)-breaking contributions
shown in eq.~(\ref{eq:dRtau}) are associated with operators of dimension two
($m^2$) and four ($\delta O_4$).

A very detailed theoretical analysis of $\delta R_\tau$ was performed
in refs.~\cite{PP:99}. The perturbative QCD corrections to the relevant
dimension-two and four operators turn out to take the values
$\Delta_{00}(\alpha_s)=2.0\pm 0.5$ and $Q_{00}(\alpha_s)= 1.08\pm 0.03$.
In order to predict $\delta R_\tau$, one needs an input value for
the strange quark mass; we will adopt the range
\begin{equation}\label{eq:ms}
m_s(m_\tau) = (100\pm 10)\:\mathrm{MeV}
 \qquad \qquad
[m_s(2\:\mathrm{GeV}) = (96\pm 10)\:\mathrm{MeV}]\, ,
\end{equation}
which includes the most recent determinations of $m_s$ from QCD sum rules
and lattice QCD \cite{JOP:06}.
This gives $\delta R_\tau = 0.227\pm 0.054$, which implies
\begin{equation}
|V_{us}| \, =\, 0.2216\pm 0.0031_{\mathrm{\, exp}}\pm 0.0017_{\mathrm{\, th}}
\, =\, 0.2216\pm 0.0036\, .
\end{equation}

The largest theoretical uncertainty originates in the longitudinal contribution
($J=L$) to the dimension-two correction $\Delta_{00}(\alpha_s)$. The corresponding
perturbative series, which is known to $O(\alpha_s^3)$, shows a very pathological
behaviour with clear signs of being non-convergent; this induces a large theoretical
error in $\Delta_{00}(\alpha_s)$. Fortunately, the longitudinal
contribution to $R_\tau$ can be estimated phenomenologically with a much higher
accuracy, because it is dominated by far by the well-known pion and kaon poles,
\begin{equation}
\frac{1}{\pi}\,\mathrm{Im}\,\Pi_{ud,A}^L(s)\, =\, 2 f_\pi^2\,\delta(s-m_\pi^2)\, ,
\qquad\qquad
\frac{1}{\pi}\,\mathrm{Im}\,\Pi_{us,A}^L(s)\, =\, 2 f_K^2\,\delta(s-m_K^2)\, ,
\end{equation}
which are determined by the $\pi^-\to l^-\bar\nu_l$ and $K^-\to l^-\bar\nu_l$ decay
widths.
Although much smaller, the leading contribution to the scalar spectral function
can be also obtained from s-wave $K\pi$ scattering data \cite{JOP:06,JOP:02}.
Taking into account additional tiny corrections from higher-mass pseudoscalar
resonances \cite{KM:02}, one
obtains the following phenomenological determination of the
longitudinal contribution to $\delta R_\tau$ \cite{GJPPS:05,JOP:06}:
\begin{equation}\label{eq:Rlon}
\delta R_\tau|^{L}\, =\, 0.1544\pm 0.0037\, .
\end{equation}
The pion and kaon contributions amount to 79\% of $\delta R_\tau|^{L}$.
For comparison, taking the strange quark mass in the range (\ref{eq:ms}),
the direct QCD calculation of this quantity
gives $\delta R_\tau|^{L} = 0.166\pm 0.051$; in agreement with (\ref{eq:Rlon}),
but with a much larger error.

The perturbative QCD series $\Delta_{00}^{L+T}(\alpha_s)$ is much better behaved,
although it starts to show an asymptotic character at $O(\alpha_s^3)$. Following
the prescription advocated in refs.~\cite{PP:99}, one finds
$\frac{3}{4}\,\Delta_{00}^{L+T}(\alpha_s)= 0.75\pm 0.12$ and
$Q_{00}^{L+T}(\alpha_s)= 0.08\pm 0.03$, which imply
$\delta R_\tau|^{L+T} = 0.062\pm 0.015$. Adding the longitudinal
contribution (\ref{eq:Rlon}), gives $\delta R_\tau = 0.216\pm 0.016$, which
agrees with the pure QCD determination and is 3.4 times more precise. This allows
us to obtain an improved $V_{us}$ determination:
\begin{equation}\label{eq:det1}
|V_{us}| \, =\, 0.2212\pm 0.0031_{\mathrm{\, exp}}\pm 0.0005_{\mathrm{\, th}}
\, =\, 0.2212\pm 0.0031\, .
\end{equation}

\section{Improved evaluation of \ $\Gamma\mathbf{\left(\tau^-\to\nu_\tau K^-/\pi^-\right)}$}

The phenomenological determination of $\delta R_\tau|^{L}$ contains a hidden
dependence on $V_{us}$ through the input value of the kaon decay constant
$f_K$. Although the numerical impact of this dependence is negligible,
it is worth while to take it explicitly into account. At the same time, we
can determine the $\tau^-\to\nu_\tau K^-/\pi^-$ decay widths with better
accuracy than the present direct experimental measurements, through the ratios
($P=K,\pi$)
\begin{equation}
R_{\tau/P}\,\equiv\,
\frac{\Gamma\left(\tau^-\to\nu_\tau P^-\right)}{\Gamma\left(P^-\to\bar\nu_\mu \mu^-\right)}
\, =\,\frac{m_\tau^3}{2m_P m_\mu^2}\,
\frac{\left( 1-m_P^2/m_\tau^2\right)^2}{\left( 1-m_\mu^2/m_P^2\right)^2}\,
\left( 1+\delta R_{\tau/P}\right)\, ,
\end{equation}
where $\delta R_{\tau/\pi}=(0.16\pm 0.14)\% $ and $\delta R_{\tau/K}=(0.90\pm 0.22)\% $
are the estimated electroweak radiative corrections \cite{MS:93,DF:95}.
Using the measured $K^-/\pi^-\to\bar\nu_\mu \mu^-$ decay widths and the $\tau$
lifetime \cite{PDG}, one gets then:
\begin{equation}\label{eq:tautoK}
\mathrm{Br}(\tau^-\to\nu_\tau K^-) = (0.715\pm 0.004)\%\, ,
\qquad\quad
\mathrm{Br}(\tau^-\to\nu_\tau \pi^-) = (10.90\pm 0.04)\%\, ,
\end{equation}
in good agreement with the less accurate PDG averages
$\mathrm{Br}(\tau^-\to\nu_\tau K^-) = (0.691\pm 0.023)\%$
and $\mathrm{Br}(\tau^-\to\nu_\tau \pi^-) = (10.90\pm 0.07)\%$.

Following ref.~\cite{DHZ:05}, we will use the improved estimate
of the electronic branching fraction
$\mathrm{Br}(\tau^-\to\nu_\tau\bar\nu_e e^-)_{\mathrm{univ}} = (17.818\pm 0.032)\%$,
which is obtained by averaging the direct measurements of the electronic an
muonic branching fractions and the $\tau$ lifetime, assuming universality.
The resulting kaon and pion contributions to $R_\tau$ take then the values:
\begin{equation}
R_\tau|^{\tau^-\to\nu_\tau K^-} = (0.04014\pm 0.00021)\, ,
\qquad\quad
R_\tau|^{\tau^-\to\nu_\tau \pi^-} = (0.6123\pm 0.0025)\, .
\end{equation}

In our theoretical analysis of $\delta R_\tau$, we have considered the separate
contributions from the $J=L$ and $J=L+T$ pieces, defined through the two terms
on the rhs of eqs.~(\ref{eq:contour}). The corresponding splitting
of the kaon and pion contributions is given by:
\begin{eqnarray}
R_\tau|^{\tau^-\to\nu_\tau K^-}_{_L} &=&\, R_\tau|^{\tau^-\to\nu_\tau K^-} -\,
R_\tau|^{\tau^-\to\nu_\tau K^-}_{_{L+T}} =\;
-2\,\frac{m_K^2}{m_\tau^2}\; R_\tau|^{\tau^-\to\nu_\tau K^-}
 =\; -(0.006196\pm 0.000033)\, ,
\nonumber\\
R_\tau|^{\tau^-\to\nu_\tau \pi^-}_{_L} &=&\, R_\tau|^{\tau^-\to\nu_\tau \pi^-} -\,
R_\tau|^{\tau^-\to\nu_\tau \pi^-}_{_{L+T}}\, =\;
-2\,\frac{m_\pi^2}{m_\tau^2}\; R_\tau|^{\tau^-\to\nu_\tau \pi^-}
\, =\; -(0.007554\pm 0.000031)\, .\nonumber\\
\end{eqnarray}

Subtracting the longitudinal contributions from eq.~(\ref{eq:Vus_formula}),
we can give an improved formula to determine $V_{us}$ with the best possible
accuracy:
\begin{equation}\label{eq:Vus_improvedForm}
 |V_{us}|^2 \, =\, \frac{\tilde R_{\tau,S}}{\frac{\tilde R_{\tau,V+A}}{|V_{ud}|^2}-\delta
 \tilde R_{\tau,\mathrm{th}}} \,\equiv\,
 \frac{R_{\tau,S} - R_\tau|^{\tau^-\to\nu_\tau K^-}_{_L}}{\frac{R_{\tau,V+A}
 -R_\tau|^{\tau^-\to\nu_\tau \pi^-}_{_L}}{|V_{ud}|^2}-
 \delta\tilde R_{\tau,\mathrm{th}}}
 \, ,
\end{equation}
where
\begin{equation}
\delta\tilde R_{\tau,\mathrm{th}}\,\equiv\, \delta\tilde R_\tau|^{L}
+ \delta R_{\tau,\mathrm{th}}|^{L+T}\, =\,
(0.033\pm 0.003) + (0.062\pm 0.015)\, =\, 0.095\pm 0.015\, .
\end{equation}
The subtracted longitudinal correction $\delta\tilde R_\tau|^{L}$ is now much
smaller because it does not contain any pion or kaon contribution.

Using the same input values for $R_{\tau,S}$ and $R_{\tau,V+A}$, one recovers
the $V_{us}$ determination obtained before in eq.~(\ref{eq:det1}), with a slightly
improved error of $\pm\, 0.0030$.

\section{Experimental update of $\mathbf{R_{\tau,S}}$}

Within the Standard Model, where charged-current universality holds,
the electron branching fraction
$B_e\equiv\mathrm{Br}(\tau^-\to\nu_\tau\bar\nu_e e^-)_{\mathrm{univ}} = (17.818\pm 0.032)\%$
determines the hadronic one, i.e.,
$R_\tau = \frac{1}{B_e}-1.972564 = 3.640\pm 0.010$.
Since $R_{\tau,V+A} = R_\tau - R_{\tau,S}$, the only additional experimental input
which is needed is $R_{\tau,S}$. Up to now, we have been using the value
$R_{\tau,S}=0.1686\pm 0.0047$, from a recent compilation of LEP and CLEO data \cite{DHZ:05}.

Babar and Belle have recently reported their first measurements of Cabibbo-suppressed
$\tau$ decays:
$\mathrm{Br}(\tau^-\to\nu_\tau\phi K^-) =(4.05\pm 0.25\pm 0.26)\cdot 10^{-5}$ \cite{BE:06},
$\mathrm{Br}(\tau^-\to\nu_\tau K_S\pi^-) =(0.404\pm 0.002\pm 0.013)\% $ \cite{BE:07},
$\mathrm{Br}(\tau^-\to\nu_\tau K^-\pi^0) =(0.416\pm 0.003\pm 0.018)\% $ \cite{BA:07a},
$\mathrm{Br}(\tau^-\to\nu_\tau K^-\pi^-\pi^+) =(0.273\pm 0.002\pm 0.009)\% $ \cite{BA:07b}
and
$\mathrm{Br}(\tau^-\to\nu_\tau\phi K^-) =(3.39\pm 0.20\pm 0.28)\cdot 10^{-5}$ \cite{BA:07b}.
The last mode includes
$\mathrm{Br}(\tau^-\to\nu_\tau K^-K^-K^+) =(1.58\pm 0.13\pm 0.12)\cdot 10^{-5}$ \cite{BA:07b},
which is found to be consistent with going entirely through $\tau^-\to\nu_\tau\phi K^-$.
The changes induced in $R_{\tau,S}$ have been nicely summarized at this workshop
by Swagato Banerjee \cite{Banerjee}. Taking for $\mathrm{Br}(\tau^-\to\nu_\tau K^-)$ the value
(\ref{eq:tautoK}) and including the small $\tau^-\to\nu_\tau\phi K^-$ decay mode, one finds
a total probability for the $\tau$ to decay into strange final states of
$(2.882\pm 0.071)\% $, which implies the updated values
\begin{equation}
R_{\tau,S}=0.1617\pm 0.0040\, ,
\qquad\qquad\qquad
R_{\tau,V+A}=3.478\pm 0.011\, .
\end{equation}

Although consistent within the quoted uncertainties, the new Babar and Belle measurements are all
smaller than the previous world averages, which translates into a smaller value of $R_{\tau,S}$
and $V_{us}$. Using eq.~(\ref{eq:Vus_improvedForm}), one finds now
\begin{equation}\label{eq:det2}
|V_{us}| \, =\, 0.2165\pm 0.0026_{\mathrm{\, exp}}\pm 0.0005_{\mathrm{\, th}}
\, =\, 0.2165\pm 0.0026\, .
\end{equation}

Sizeable changes on the experimental determination of $R_{\tau,S}$ are to be expected from
the full analysis of  the huge Babar and Belle data samples. In particular, the high-multiplicity
decay modes are not well known at present and their effect has been just roughly estimated or
simply ignored. Thus, the result (\ref{eq:det2}) could easily fluctuate in the near future.
However, it is important to realize that the final error of the $V_{us}$ determination from
$\tau$ decay is completely dominated by the experimental uncertainties. If $R_{\tau,S}$
is measured with a 1\% precision, the resulting $V_{us}$ uncertainty will
get reduced to around 0.6\%, i.e. $\pm 0.0013$, making $\tau$ decay the best source of
information about $V_{us}$.

An accurate experimental measurement of the invariant-mass distribution of the final
hadrons in Cabibbo-suppressed $\tau$ decays could make possible a simultaneous determination
of $V_{us}$ and the strange quark mass. However, the extraction of $m_s$ suffers from
theoretical uncertainties related to the convergence of the perturbative series
$\Delta_{00}^{L+T}(\alpha_s)$. A better understanding of these corrections is needed
\cite{GJPPS:07}.

\acknowledgments

This work has been supported by
the European Commission MRTN FLAVIA{\it net}
[MRTN-CT-2006-035482], the MEC (Spain) and FEDER (EC)
[FPA2005-02211 (M.J.), FPA2004-00996 (A.P.) and FPA2006-05294 (J.P.)],
the Deutsche Forschungsgemeinschaft (F.S.),
the Junta de Andaluc\'\i a [P05-FQM-101 (J.P.), P05-FQM-437 (E.G. and J.P.)
and Sabbatical Grant PR2006-0369 (J.P.)] and
the Generalitat Valenciana [GVACOMP2007-156 (A.P.)].

\end{document}